\documentclass{jps-cp}
\usepackage{txfonts}

\usepackage{color}

\def\gsim{\mathop {\vtop {\ialign {##\crcr 
$\hfil \displaystyle {>}\hfil $\crcr \noalign {\kern1pt \nointerlineskip } 
$\,\sim$ \crcr \noalign {\kern1pt}}}}\limits}
\def\lsim{\mathop {\vtop {\ialign {##\crcr 
$\hfil \displaystyle {<}\hfil $\crcr \noalign {\kern1pt \nointerlineskip } 
$\,\,\sim$ \crcr \noalign {\kern1pt}}}}\limits}

\title{
Spin-orbit-phonon interaction as an origin of helical-symmetry breaking spin-triplet superconducting state
}

\author{
Kazumasa Miyake
}
\inst{
Center for Advanced High Magnetic Field Science, Osaka University, Toyonaka, 560-0043, Japan 
}

\email{
miyake@mp.es.osaka-u.ac.jp}

\recdate{September 17, 2019}

\abst{
The excess increase of  the local field distribution width 
$\Delta H\equiv\sigma/\gamma_{\mu}$ (with $\gamma_{\mu}=135.5\,$MHz/T being the gyromagnetic ratio 
of $\mu^{+}$), and/or the spontaneous magnetic field $H_{\rm spon}$, have been reported  
in the superconducting state of a series of compounds more than 10 
by $\mu$SR (muon spin rotation or relaxation) experiment.  
Sizes of these quantities are of  the order of 1G on the whole. 
We propose that the increase of the local field distribution $\Delta H$ in those compounds, including ions 
with strong spin-orbit interaction of 5d electrons, such as La in LaNiGa$_2$ or Re in Re$_6$Zr, is possible 
through the cooperation of the spin-orbit coupling between the quasiparticles and 
ionic vibrations (phonons) and the conventional quasiparticles-phonon coupling, 
inducing the spin-triplet p-wave pairing with helical symmetry breaking superconducting state 
with chiral super spin current around stopped $\mu^{+}$ site. The size of $\Delta H$ is estimated as   
of the order of 1G, in consistent with experimental observations.  
}

\begin{document}
\maketitle

\section{Introduction}
In a past decade or so, the extra increase of the {\it local field distribution} (LFD) in the superconducting state 
has been observed 
by the muon spin rotaion ($\mu$SR) measurements in a series of 
compounds~\cite{Ghosh}.
In these measurements, the time dependence of $\mu$SR spectrum was analyzed by fitting to 
the sinusoidal oscillating function with Gaussian relaxation $\propto \exp(-\sigma^{2}t^{2}/2)$,
from which the LFD is estimated as $\Delta H\equiv \sigma/\gamma_{\mu}$ with the muon 
gyromagnetic ratio $\gamma_{\mu}=2\pi\times 135.5$ MHz$\,T^{-1}$~\cite{Hillier2012}. 
The sizes of the LFD  
so determined are all of the order of 1G on the whole, suggesting that 
there exists a common physical basis for the increase of LFD in the superconducting states. 

On the other hand, 
the {\it spontaneous} magnetic field $H_{\rm spon}$ of the order of 1G 
has been observed by $\mu$SR in some superconductors, 
e.g., Sr$_2$RuO$_4$~\cite{Luke} which is believed to be in the {\it unitary} 
spin-triplet p-wave chiral superconducting state 
$\Delta_{\bf k}=\Delta(\sin k_{x}a+{\rm i}\sin k_{y}a)$ with $a$ being the lattice constant in 
the $ab$-plane~\cite{Miyake2, Mackenzie}. 
This phenomenon was shown to be understood as a pair-breaking effect of $\mu^{+}$ which attracts 
electron on the Ru site and breaks locally the chiral Cooper pairs around $\mu^{+}$, 
causing the circulating super current around $\mu^{+}$ resulting in $H_{\rm spon}$ on $\mu^{+}$ 
{\it in turn}~\cite{Miyake3}.  
Namely, it is crucial to realize that the muon is not only the probe measuring properties of the system 
but changes also the local physical property of the system. 
 


The purpose of this paper is to propose a new mechanism for the extra 
LFD $\Delta H$ 
on the basis of a helical-symmetry breaking spin-triplet p-wave superconducting state 
which can give rise to the spontaneous super spin current around stopped $\mu^{+}$ 
leading to the increase 
in the LFD $\Delta H$ in the superconducting state through the spin-flipping of $\mu^{+}$ 
caused by the dipole interaction between the spin of the Cooper pairs circulating around the muon. 
Such a superconducting state is shown to be possible by a cooperation of 
strong spin-orbit-phonon coupling and electron-phonon interaction.  



\section{Pairing Interaction Induced by Spin-Orbit-Phonon and Electron-Phonon Interactions} 
\subsection{Spin-orbit-phonon interaction}
The scattering vertex $\Gamma^{\rm so}_{{\bf k},\sigma;{\bf k}^{\prime},\sigma^{\prime}}({\bf R}_{n})$ 
of conduction electrons from $({\bf k},\sigma)$ to $({\bf k}^{\prime},\sigma^{\prime})$ 
by spin-orbit interaction from the atom located at ${\bf R}_n$ is given as 
\begin{eqnarray}
\Gamma^{\rm so}_{{\bf k},\sigma;{\bf k}^{\prime},\sigma^{\prime}}({\bf R}_{n})={-}
{\rm i}g\sum_{\sigma=\pm}\sum_{\sigma^{\prime}=\pm}\sum_{{\bf k},{\bf k}^{\prime}}
[{\bf s}_{\sigma\sigma^{\prime}}\cdot({\bf k}\times{\bf k}^{\prime})]
U_{{\bf k}-{\bf k}^{\prime}}({\bf R}_{n})c^{\dagger}_{{\bf k}\sigma}c_{{\bf k}^{\prime}\sigma^{\prime}},
\label{SO1}
\end{eqnarray}
where {$g\approx e\hbar^{2}/4mm^{*}c^{2}>0$ with $m^{*}$ being the effective mass 
of quasiparticles, and} 
${\bf s}$ is the spin operator {in the unit of $\hbar$}, 
and $U_{{\bf k}-{\bf k}^{\prime}}({\bf R}_{n})$ is 
defined in terms of the atomic potential located at 
${\bf R}_{n}$ as 
\begin{eqnarray}
& &
U_{{\bf k}-{\bf k}^{\prime}}({\bf R}_{n})\equiv
\int {\rm d}{\bf r}e^{-{\rm i}({\bf k}-{\bf k}^{\prime})\cdot {\bf r}}U({\bf r}-{\bf R}_{n})
=e^{-{\rm i}({\bf k}-{\bf k}^{\prime})\cdot {\bf R}_{n}}U_{{\bf k}-{\bf k}^{\prime}},
\label{SO2}
\end{eqnarray}
where $U_{\bf q}\equiv\int {\rm d}{\bf r}e^{-{\rm i}{\bf q}\cdot{\bf r}}U({\bf r})$. 
Hereafter we assume that the conduction electrons are described essentially by the free dispersion and 
the spin degrees of freedom.  
It is crucial to note that such a strong spin-orbit interaction can be induced through the 
hybridization between conduction electrons and the electrons in the atomic orbitals which are subject 
to the strong spin-orbit interaction from the positive nuclear charge in heavy ions such as Re, 
while the direct screened Coulomb interaction from the nuclear charge is far less important.    

Since ${\bf R}_{n}$ oscillates by the influence of the phonon vibrations, the position of the atom is 
expressed as ${\bf R}_{n}={\bar{\bf R}_{n}}+{\bf u}_{n}$ where ${\bar{\bf R}_{n}}$ and ${\bf u}_{n}$ are 
the equilibrium position of $n$-teh atom and the deviation from it, respectively. 
Then, by taking the summation with respect to ${\bar {\bf R}}_{n}$, 
the interaction $\Gamma^{\rm so-ph}$ between the spin-orbit interaction and  
phonon vibrations is given  as 
\begin{eqnarray}
\Gamma^{\rm so-ph}={-}{\rm i}g\sum_{\sigma,\sigma^{\prime}}\sum_{{\bf k},{\bf k}^{\prime}}
[{\bf s}_{\sigma\sigma^{\prime}}\cdot({\bf k}\times{\bf k}^{\prime})]
(-{\rm i})({\bf k}-{\bf k}^{\prime})\cdot{\bf u}_{{\bf k}-{\bf k}^{\prime}}
{U_{{\bf  k}-{\bf k}^{\prime}}}
c^{\dagger}_{{\bf k}\sigma}c_{{\bf k}^{\prime}\sigma^{\prime}},
\label{SO6}
\end{eqnarray}
where 
${\bf u}_{{\bf q}}\equiv\sum_{n}{\bf u}_{n}e^{-{\rm i}{\bf q}\cdot {\bar {\bf R}}_{n}}$.
Since ${\rm i}{\bf q}\cdot{\bf u}_{\bf q}$ is described by the phonon creation and 
annihilation operaters as $A_{\bf q}(b_{\bf q}+b^{\dagger}_{-{\bf q}})$ with ${\bf q}$ dependent coefficient 
$A_{\bf q}$~\cite{AGD}, the spin-orbit coupling and phonon interaction is given by a simple form as 
\begin{eqnarray}
\Gamma^{\rm so-ph}={\rm i}g\sum_{\sigma,\sigma^{\prime}}\sum_{{\bf k},{\bf k}^{\prime}}
[{\bf s}_{\sigma\sigma^{\prime}}\cdot({\bf k}\times{\bf k}^{\prime})]
{U_{{\bf  k}-{\bf k}^{\prime}}}
A_{{\bf k}-{\bf k}^{\prime}}\left(b_{{\bf k}-{\bf k}^{\prime}}+b^{\dagger}_{-{\bf k}+{\bf k}^{\prime}}\right)
c^{\dagger}_{{\bf k}\sigma}c_{{\bf k}^{\prime}\sigma^{\prime}}. 
\label{SO7}
\end{eqnarray}

\subsection{Quasiparticle-phonon interaction }
The quasiaprticle-phonon interaction $\Gamma^{\rm el-ph}$ is represented as 
\begin{eqnarray}
\Gamma^{\rm el-ph}=\sum_{\sigma}\sum_{{\bf k},{\bf k}^{\prime}}
W_{{\bf k},{\bf k}^{\prime}}
A_{{\bf k}-{\bf k}^{\prime}}\left(b_{{\bf k}-{\bf k}^{\prime}}+b^{\dagger}_{-{\bf k}+{\bf k}^{\prime}}\right)
c^{\dagger}_{{\bf k}\sigma}c_{{\bf k}^{\prime}\sigma},
\label{SO8}
\end{eqnarray}
where the structure factor $W_{{\bf k},{\bf k}^{\prime}}$ depends on the origin of the electron-phonon 
interaction. Namely, in the free electron picture where the coupling arises from the ionic charge 
accumulation $(-e)[-{\rm div}{\bf u}({\bf r})]$ at ${\bf r}={\bf R}_{n}$ which influences the electrons through 
the screened Coulomb potential~\cite{AGD}, while  
in the tight-binding picture, variation of ions does not break charge neutrality 
associated with motion of ions 
so that the electron-phonon  coupling arises through the variation of the transfer integral corresponding 
to the variation of distance among ions due to the lattice vibrations~\cite{Miyake4}.   

\subsection{General expression of pairing interaction triggered by spin-orbit-phonon and 
electron-phonon interactions}
\begin{figure}[h]
\begin{center}
\rotatebox{0}{\includegraphics[width=0.7\linewidth]{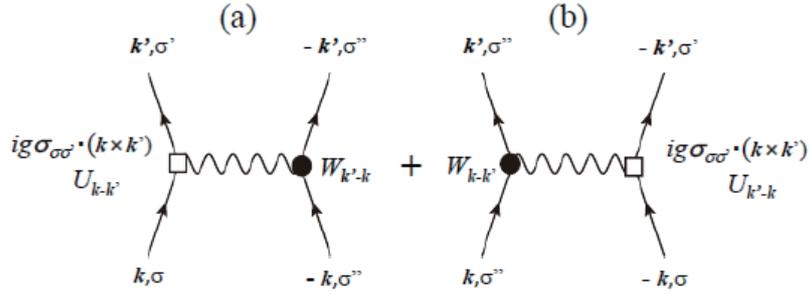}}
\caption{Feynman diagram for the pairing interaction induced by by the spin-orbit-phonon and 
electron-phonon  interactions. The open square and filled circle represent the spin-orbit-phonon  
coupling and the conventional electron-phonon coupling, respectively. }
\label{Fig:1}
\end{center}
\end{figure}

Spin-orbit coupling and phonon interaction and electron phonon interaction induces the pairing 
interaction $V^{\rm so-ph}_{{\bf k}-{\bf k}^{\prime}}$ by the Feynman diagram shown in Fig.\ \ref{Fig:1}.  
Explicit form of $V^{\rm so}_{{\bf k}-{\bf k}^{\prime}}$ in the static limit is given as 
follows: 
\begin{eqnarray}
& &
V^{\rm so-ph}_{{\bf k}-{\bf k}^{\prime},\sigma\sigma^{\prime}}
={\rm i}g[{\bf s}_{\sigma\sigma^{\prime}}\cdot
({\bf k}\times{\bf k}^{\prime})]U_{{\bf k}-{\bf k}^{\prime}}A_{{\bf k}^{\prime}-{\bf k}}^{ 2}
W_{{\bf k}-{\bf k}^{\prime}}D_{\rm ph}({\bf k}-{\bf k}^{\prime}, 0)
\nonumber
\\
& &
\qquad\qquad\,\,
+
{\rm i}g[{\bf s}_{\sigma\sigma^{\prime}}\cdot({\bf k}\times{\bf k}^{\prime})]
U_{{\bf k}^{\prime}-{\bf k}}A_{{\bf k}^{\prime}-{\bf k}}^{ 2}
W_{{\bf k}^{\prime}-{\bf k}}D_{\rm ph}({\bf k}^{\prime}-{\bf k}, 0), 
\label{TRBSC1}
\end{eqnarray}
where $D_{\rm ph}({\bf k}-{\bf k}^{\prime}, {\rm i}\epsilon_{n}-{\rm i}\epsilon_{n^{\prime}})$ 
is the Matsubara Green function of phonons with 
${\rm i}\epsilon_{n}$'s being the fermionic Matsubara frequencies.  
Since the wave vector 
dependence of $U_{\pm{\bf k}\mp{\bf k}^{\prime}}A_{\pm{\bf k}^{\prime}\mp{\bf k}}^{ 2}
W_{\pm{\bf k}\mp{\bf k}^{\prime}}D_{\rm ph}(\pm{\bf k}\mp{\bf k}^{\prime}, 0)$  
{
is expected to be week and negative,}
according to the fact that the conventional phonon mediated Cooper pair is the s-wave which is 
essentially wave vector independent. Then, denoting it by $-\Lambda$, the paring interaction 
[Eq.\ (\ref{TRBSC1})] is reduced to a simple form as 
\begin{eqnarray}
& &
V^{\rm so-ph}_{{\bf k}-{\bf k}^{\prime},\sigma\sigma^{\prime}}
=-2{\rm i}g\Lambda
[{\bf s}_{\sigma\sigma^{\prime}}\cdot({\bf k}\times{\bf k}^{\prime})]. 
\label{TRBSC1A}
\end{eqnarray}

\section{Helical-Symmetry Breaking Cooper Pairs}
The pairing interaction [Eq.\ (\ref{TRBSC1A})] has different wave vector dependence depending 
on the existence or non-existence of spin flipping process. Namely, in the case without 
spin flip ($\sigma^{\prime}=\sigma=\pm$), 
\begin{eqnarray}
& &
V^{\rm so-ph\,\sigma^{\prime\prime}}_{{\bf k}-{\bf k}^{\prime},\sigma\sigma}
=-{\rm i}g\Lambda\sigma(k_{x}k_{y}^{\prime}-k_{y}k_{x}^{\prime}),
\label{TRBSC2}
\end{eqnarray}
while in the case with spin flip ($\sigma^{\prime}=-\sigma$), 
\begin{eqnarray}
& &
V^{\rm so-ph\,\sigma^{\prime\prime}}_{{\bf k}-{\bf k}^{\prime},+-}
=-g\Lambda\left[-(k_{x}-{\rm i}k_{y})k_{z}^{\prime}+k_{z}(k_{x}^{\prime}-{\rm i}k_{y}^{\prime})\right],
\label{TRBSC3+-}
\end{eqnarray}
or 
\begin{eqnarray}
& &
V^{\rm so-ph\,\sigma^{\prime\prime}}_{{\bf k}-{\bf k}^{\prime},-+}
=-g\Lambda\left[(k_{x}+{\rm i}k_{y})k_{z}^{\prime}-k_{z}(k_{x}^{\prime}+{\rm i}k_{y}^{\prime})\right].
\label{TRBSC3-+}
\end{eqnarray}
Note that paring interactions [Eqs.\ (\ref{TRBSC2})-(\ref{TRBSC3-+})] induce triplet p-wave pairings 
in one form or another,   
and are independent of $\sigma^{\prime\prime}$,  
the spin component of electrons scattered by phonons. 

Hereafter, we choose the $z$-direction as a special and favorable one for the Cooper pair formation 
on the basis of the assumption that 
the effective mass of the quasipartices in the $xy$-plane is considerably larger than that in the 
$z$-direction, which makes the spin-orbit interaction for $S_{z}$ dominant compared to 
those for $S_{x}$ and $S_{y}$.   

\subsection{Gap equations at transition temperature}
Gap symmetry at the transition temperature$T_{\rm c}$  is determined by the gap equation 
with the pairing interactions [Eqs.\ (\ref{TRBSC2})-(\ref{TRBSC3-+})]. 
Considering these interactions induce 
the spin-triplet p-wave pairing, let us introduces the four gap functions 
$\Delta_{\sigma\sigma^{\prime}}({\bf k})$ with  ($\sigma,\,\sigma^{\prime}=+\,{\rm or}\, -$). 
Then, the gap equations at $T=T_{\rm c}$ are given by two coupled equations as follows:  
\begin{eqnarray}
& &
\Delta_{++}({\bf k})=-\sum_{{\bf k}^{\prime}}
\left[
V^{\rm so-ph\,+}_{{\bf k}-{\bf k}^{\prime},++}\Delta_{++}({\bf k}^{\prime})
+
V^{\rm so-ph\,+}_{{\bf k}-{\bf k}^{\prime},+-}\Delta_{-+}({\bf k}^{\prime})\right]
\Phi_{{\bf k}^{\prime}}
\label{Gap++}
\\
& &
\Delta_{-+}({\bf k})=-\sum_{{\bf k}^{\prime}}
\left[
V^{\rm so-ph\,+}_{{\bf k}-{\bf k}^{\prime},--}\Delta_{-+}({\bf k}^{\prime})
+
V^{\rm so-ph\,+}_{{\bf k}-{\bf k}^{\prime},-+}\Delta_{++}({\bf k}^{\prime})\right]
\Phi_{{\bf k}^{\prime}}
\label{Gap-+}
\end{eqnarray}
and 
\begin{eqnarray}
& &
\Delta_{--}({\bf k})=-\sum_{{\bf k}^{\prime}}
\left[
V^{\rm so-ph\,-}_{{\bf k}-{\bf k}^{\prime},--}\Delta_{--}({\bf k}^{\prime})
+
V^{\rm so-ph\,-}_{{\bf k}-{\bf k}^{\prime},-+}\Delta_{+-}({\bf k}^{\prime})\right]
\Phi_{{\bf k}^{\prime}}
\label{Gap--}
\\
& &
\Delta_{+-}({\bf k})=-\sum_{{\bf k}^{\prime}}
\left[
V^{\rm so-ph\,-}_{{\bf k}-{\bf k}^{\prime},++}\Delta_{+-}({\bf k}^{\prime})
+
V^{\rm so-ph\,-}_{{\bf k}-{\bf k}^{\prime},+-}\Delta_{--}({\bf k}^{\prime})\right]
\Phi_{{\bf k}^{\prime}}
\label{Gap+-},
\end{eqnarray}
where $\Phi_{{\bf k}^{\prime}}\equiv {\rm tanh}(\xi_{{\bf k}^{\prime}}/2T_{\rm c})/(2\xi_{{\bf k}^{\prime}})$ 
with the dispersion $\xi_{{\bf k}^{\prime}}$ of the quasiparticles.  

With the use of the expressions [Eqs.\ (\ref{TRBSC2})-(\ref{TRBSC3-+})] for the pairing interactions, 
Eqs.\ (\ref{Gap++}) and (\ref{Gap-+}) are given explicitly as 
\begin{eqnarray}
& &
\Delta_{++}({\bf k})={\rm i}g\Lambda\sum_{{\bf k}^{\prime}}
(k_{x}k^{\prime}_{y}-k_{y}k^{\prime}_{x})\Delta_{++}({\bf k}^{\prime})\Phi_{{\bf k}^{\prime}}
\nonumber
\\
& &
\qquad\qquad\quad
+g\Lambda\sum_{{\bf k}^{\prime}}
[-(k_{x}-{\rm i}k_{y})k_{z}^{\prime}+k_{z}(k_{x}^{\prime}-{\rm i}k_{y}^{\prime})]
\Delta_{-+}({\bf k}^{\prime})\Phi_{{\bf k}^{\prime}},
\label{GapExplicit++}
\\
& &
\Delta_{-+}({\bf k})=-{\rm i}g\Lambda\sum_{{\bf k}^{\prime}}
(k_{x}k^{\prime}_{y}-k_{y}k^{\prime}_{x})\Delta_{-+}({\bf k}^{\prime})\Phi_{{\bf k}^{\prime}}
\nonumber
\\
& &
\qquad\qquad\quad
+g\Lambda\sum_{{\bf k}^{\prime}}
[(k_{x}+{\rm i}k_{y})k_{z}^{\prime}-k_{z}(k_{x}^{\prime}+{\rm i}k_{y}^{\prime})]
\Delta_{++}({\bf k}^{\prime})\Phi_{{\bf k}^{\prime}}.
\label{GapExplicit-+}
\end{eqnarray}
Similarly, Eqs.\ (\ref{Gap--}) and (\ref{Gap+-}) are given explicitly as 
\begin{eqnarray}
& &
\Delta_{--}({\bf k})=-{\rm i}g\Lambda\sum_{{\bf k}^{\prime}}
(k_{x}k^{\prime}_{y}-k_{y}k^{\prime}_{x})\Delta_{--}({\bf k}^{\prime})\Phi_{{\bf k}^{\prime}}
\nonumber
\\
& &
\qquad\qquad\quad
+g\Lambda\sum_{{\bf k}^{\prime}}
[(k_{x}+{\rm i}k_{y})k_{z}^{\prime}-k_{z}(k_{x}^{\prime}+{\rm i}k_{y}^{\prime})]
\Delta_{+-}({\bf k}^{\prime})\Phi_{{\bf k}^{\prime}},
\label{GapExplicit--}
\\
& &
\Delta_{+-}({\bf k})={\rm i}g\Lambda\sum_{{\bf k}^{\prime}}
(k_{x}k^{\prime}_{y}-k_{y}k^{\prime}_{x})\Delta_{+-}({\bf k}^{\prime})\Phi_{{\bf k}^{\prime}}
\nonumber
\\
& &
\qquad\qquad\quad
+g\Lambda\sum_{{\bf k}^{\prime}}
[-(k_{x}-{\rm i}k_{y})k_{z}^{\prime}+k_{z}(k_{x}^{\prime}+{\rm i}k_{y}^{\prime})]
\Delta_{--}({\bf k}^{\prime})\Phi_{{\bf k}^{\prime}}.
\label{GapExplicit+-}
\end{eqnarray}


In order to find possible gap symmetries satisfying Eqs.\ (\ref{GapExplicit++})-(\ref{GapExplicit+-}), 
we postulate the ${\bf k}$ dependence of gap functions as follows:
\begin{eqnarray}
& &
\Delta_{++}({\bf k})={\bar \Delta}[a_{+}(k_{x}-{\rm i}k_{y})+b_{+}(k_{x}+{\rm i}k_{y})+c_{+}k_{z}],
\label{Delta++}
\\
& &
\Delta_{-+}({\bf k})=
{\bar \Delta}[a_{\rm od}(k_{x}-{\rm i}k_{y})+b_{\rm od}(k_{x}+{\rm i}k_{y})+c_{\rm od}k_{z}],
\label{Delta-+}
\\
& &
\Delta_{--}({\bf k})=
{\bar \Delta}[a_{-}(k_{x}-{\rm i}k_{y})+b_{-}(k_{x}+{\rm i}k_{y})+c_{-}k_{z}],
\label{Delta--}
\\
& &
\Delta_{+-}({\bf k})=
{\bar \Delta}[a_{\rm od}(k_{x}-{\rm i}k_{y})+b_{\rm od}(k_{x}+{\rm i}k_{y})+c_{\rm od}k_{z}].
\label{Delta+-}
\end{eqnarray}
Note that the two off-diagonal gaps should be the same, i.e., $\Delta_{-+}({\bf k})=\Delta_{-+}({\bf k})$, 
by the symmetry requirement in the manifold of spin-triplet odd-parity pairings. 

Hereafter, to grasp the essence of the paring mechanism triggered by the spin-orbit interaction, 
we assume the dispersion of the quasiparticles is even function of $k_{z}$, or the inversion symmetry 
is preserved in the $z$-direction. Then, Eqs. (\ref{GapExplicit++}) and (\ref{GapExplicit-+}) are 
reduced to 
\begin{eqnarray}
& &
a_{+}(k_{x}-{\rm i}k_{y})+b_{+}(k_{x}+{\rm i}k_{y})+c_{+}k_{z}=
\nonumber
\\
& &
\qquad
g\Lambda\bigl[a_{+}(k_{x}-{\rm i}k_{y})F_{+}+a_{+}(k_{x}+{\rm i}k_{y})F_{-}
-b_{+}(k_{x}+{\rm i}k_{y})F_{+}-b_{+}(k_{x}-{\rm i}k_{y})F_{-}
\nonumber
\\
& &
\qquad\qquad\quad
+2a_{\rm od}k_{z}(F_{+}-{\rm i}F_{xy})+2b_{\rm od}k_{z}F_{+}-c_{\rm od}(k_{x}-{\rm i}k_{y})F_{z}\bigr],
\label{GapDelta++}
\end{eqnarray}
and 
\begin{eqnarray}
& &
a_{\rm od}(k_{x}-{\rm i}k_{y})+b_{\rm od}(k_{x}+{\rm i}k_{y})+c_{\rm od}k_{z}=
\nonumber
\\
& &
\qquad
g\Lambda\bigl[-a_{\rm od}(k_{x}-{\rm i}k_{y})F_{+}-a_{\rm od}(k_{x}+{\rm i}k_{y})F_{-}
+b_{\rm od}(k_{x}+{\rm i}k_{y})F_{+}+b_{\rm od}(k_{x}-{\rm i}k_{y})F_{-}
\nonumber
\\
& &
\qquad\qquad\quad
-2a_{+}k_{z}(F_{+}-{\rm i}F_{xy})-2b_{+}k_{z}F_{+}+c_{+}(k_{x}+{\rm i}k_{y})F_{z}\bigr],
\label{GapDelta-+}
\end{eqnarray}
respectively. In deriving these equations, we have used the relations 
\begin{eqnarray}
\sum_{{\bf k}^{\prime}}(k_{x}k^{\prime}_{y}-k_{y}k^{\prime}_{x})
(k_{x}^{\prime}{\mp}{\rm i}k_{y}^{\prime})\Phi_{{\bf k}^{\prime}}
={\mp}{\rm i}[(k_{x}{\mp}{\rm i}k_{y})F_{+}+(k_{x}{\pm}{\rm i}k_{y})F_{-}],
\label{Supplement1}
\end{eqnarray}
where  
$F_{\pm}\equiv
({1}/{2})\sum_{{\bf k}}[k_{y}^{2}{\pm}k_{x}^{2}]\Phi_{\bf k}$, 
$F_{z}\equiv
\sum_{{\bf k}}k_{z}^{2}\Phi_{\bf k}$, 
and 
$F_{xy}\equiv\sum_{{\bf k}}k_{x}k_{y}\Phi_{\bf k}$.

Similarly, Eqs. (\ref{GapExplicit--}) and (\ref{GapExplicit+-}) are reduced to 
\begin{eqnarray}
& &
a_{-}(k_{x}-{\rm i}k_{y})+b_{-}(k_{x}+{\rm i}k_{y})+c_{-}k_{z}=
\nonumber
\\
& &
\qquad
g\Lambda\bigl[-a_{-}(k_{x}-{\rm i}k_{y})F_{+}-a_{-}(k_{x}+{\rm i}k_{y})F_{-}
+b_{-}(k_{x}+{\rm i}k_{y})F_{+}+b_{+}(k_{x}-{\rm i}k_{y})F_{-}
\nonumber
\\
& &
\qquad\qquad\quad
-2a_{\rm od}k_{z}F_{+}-2b_{\rm od}k_{z}(F_{+}+{\rm i}F_{xy})+c_{\rm od}(k_{x}+{\rm i}k_{y})F_{z}\bigr],
\label{GapDelta--}
\end{eqnarray}
and 
\begin{eqnarray}
& &
a_{\rm od}(k_{x}-{\rm i}k_{y})+b_{\rm od}(k_{x}+{\rm i}k_{y})+c_{\rm od}k_{z}=
\nonumber
\\
& &
\qquad
g\Lambda\bigl[a_{\rm od}(k_{x}-{\rm i}k_{y})F_{+}+a_{\rm od}(k_{x}+{\rm i}k_{y})F_{-}
-b_{\rm od}(k_{x}+{\rm i}k_{y})F_{+}-b_{\rm od}(k_{x}-{\rm i}k_{y})F_{-}
\nonumber
\\
& &
\qquad\qquad\quad
+2a_{-}k_{z}(F_{+}-{\rm i}F_{xy})+2b_{-}k_{z}F_{+}-c_{-}(k_{x}-{\rm i}k_{y})F_{z}\bigr].
\label{GapDelta+-}
\end{eqnarray}

\subsection{Helical-symmetry broken pairing}
Since coupled equations Eqs.\ (\ref{GapDelta++}), (\ref{GapDelta-+}), (\ref{GapDelta--}) and 
(\ref{GapDelta+-}) 
are still complicated, we make further simplification by assuming the dispersion of the quasiparticles 
satisfies the mirror symmetry concerning $xz$, $yz$ and $(x-y)z$ planes, resulting in 
$F_{xy}=F_{-}=0$, 
Then, we obtain the relations among coefficients in a compact form. Indeed, 
from a set of equations [Eqs.\ (\ref{GapDelta++}) and (\ref{GapDelta-+})], we obtain 
\begin{eqnarray}
& &
a_{+}=g\Lambda(a_{+}F_{+}-c_{\rm od}F_{z}),
\label{Coefficient1A}
\\
& &
b_{+}=-g\Lambda F_{+}b_{+},
\label{Coefficient2A}
\\
& &
c_{+}=2g\Lambda(a_{\rm od}+b_{\rm od})F_{+},
\label{Coefficient3A}
\\
& &
a_{\rm od}=-g\Lambda F_{+} a_{\rm od},
\label{Coefficient4A}
\\
& &
b_{\rm od}=g\Lambda(b_{\rm od}F_{+}+c_{+}F_{z}),
\label{Coefficient5A}
\\
& &
c_{\rm od}=-2g\Lambda(a_{+}+b_{+})F_{+}.
\label{Coefficient6A}
\end{eqnarray}
Similarly, from a set of equations [Eq.\ (\ref{GapDelta--}) and Eq.\ (\ref{GapDelta+-})], we obtain 
\begin{eqnarray}
& &
a_{-}=-g\Lambda F_{+}a_{-},
\label{Coefficient1B}
\\
& &
b_{-}=g\Lambda (b_{-}F_{+}+c_{\rm od}F_{z}),
\label{Coefficient2B}
\\
& &
c_{-}=-2g\Lambda(a_{\rm od}+b_{\rm od})F_{+},
\label{Coefficient3B}
\\
& &
a_{\rm od}=g\Lambda(a_{\rm od}F_{+}-c_{\rm od}F_{z}),
\label{Coefficient4B}
\\
& &
b_{\rm od}=-g\Lambda F_{+}b_{\rm od},
\label{Coefficient5B}
\\
& &
c_{\rm od}=2g\Lambda(a_{-}+b_{-})F_{+}.
\label{Coefficient6B}
\end{eqnarray}
From Eqs.\ (\ref{Coefficient2A}), (\ref{Coefficient4A}), (\ref{Coefficient1B}) and 
(\ref{Coefficient5B}), it is obvious that $b_{+}=a_{\rm od}=a_{-}=b_{\rm od}=0$, resulting in 
$c_{+}=c_{-}=0$ through the relations [Eqs.\ (\ref{Coefficient3A}) and (\ref{Coefficient3B})]. 
Therefore, $\Delta_{++}$ [Eq.\ (\ref{Delta++})], $\Delta_{--}$ [Eq.\ (\ref{Delta--})], and 
$\Delta_{-+}$ [Eq.\ (\ref{Delta-+})], $\Delta_{+-}$ [Eq.\ \ref{Delta+-})] are given as 
\begin{eqnarray}
& &
\Delta_{++}({\bf k})={\bar \Delta}a_{+}(k_{x}-{\rm i}k_{y}),
\label{Delta++F}
\\
& &
\Delta_{-+}({\bf k})=0, 
\label{Delta-+F}
\end{eqnarray}
and
\begin{eqnarray}
& &
\Delta_{--}({\bf k})={\bar \Delta}b_{-}(k_{x}+{\rm i}k_{y}),
\label{Delta--F}
\\
& &
\Delta_{+-}({\bf k})=0, 
\label{Delta-+F}
\end{eqnarray}
respectively. This superconducting gap matrix ${\hat \Delta}({\bf k})$ represents a sort of 
helical-symmetry broken state 
in the sense that the direction of the circulating spin current carried by the Cooper pairs 
is in the $xy$-plane and fixed in the clockwise when looking from positive side of the $z$-axis. 
This state is similar to the BW state in the sense that the Cooper pairs with 
$\uparrow\uparrow$ and $\downarrow\downarrow$ components have opposite circulations 
resulting in the spin current near the system boundary (wall of the vessel in the case of superfluid 
$^{3}$He) in principle. 
Indeed, the gap $\Delta_{\uparrow\uparrow}$ and $\Delta_{\downarrow\downarrow}$ in the BW state 
are reproduced by taking $a_{+}=-b_{-}$ in Eqs.\ (\ref{Delta++F}) and (\ref{Delta--F}).  
Of course we need to take into account the effect of higher order terms in $\Delta$'s in the 
Landau-Ginzburg free energy expansion, which is out of scope of the present note. .  

.

The transition temperature $T_{\rm c}$ is given by solving the coupled equation 
[Eqs.\ (\ref{Coefficient1A}) and (\ref{Coefficient6A}); 
Eqs.\ (\ref{Coefficient2B}) and (\ref{Coefficient6B}] as 
\begin{eqnarray}
1=g\Lambda F_{+}(T_{\rm c})\left[1+g\Lambda F_{z}(T_{\rm c})\right].
\label{Tc1}
\end{eqnarray}

\section{{Extra Local Field Distribution $\Delta H$ 
on $\mu^{+}$ by Cooper-Pair Spin Current 
through Magnetic Dipole Interaction among the Muon and the Cooper Pairs}}  

\begin{figure}[h]
\begin{center}
\rotatebox{0}{\includegraphics[width=0.5\linewidth]{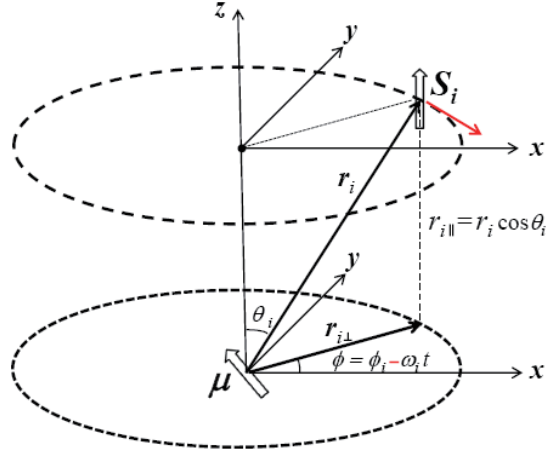}}
\caption{(Color online) 
Schematic picture of relative position of stopped $\mu^{+}$ and the spin super 
current (indicated by red arrow) induced by the Cooper pairs at ${\bf r}_{i}$ destroyed by the $\mu^{+}$ itself, 
Typical distance $r_{i}$ is of the order of the size of the Cooper pairs 
$\xi_{0}$, i.e., $r_{i}\sim \xi$. }
\label{Fig:dipole}
\end{center}
\end{figure}

The dipole-dipole interaction $H_{\rm d}$ between the muon spin ${\bf \mu}$ and electron spin  
${\bf S}_{i}$ at ${\bf r}_{i}$ is given by 
\begin{eqnarray}
& &
H_{\rm d}={\mu_{\rm B}}^{2}\gamma_{\mu}\gamma_{\rm e}
\frac{({\bf \mu}\cdot{\bf S}_{i})r_{i}^{2}-3({\bf \mu}\cdot{{\bf r}_{i}})({\bf S}_{i}\cdot{{\bf r}_{i}})}{r_{i}^{5}},
\label{dipole1}
\end{eqnarray}
where $\mu_{\rm B}$ is the Bohr magneton, $\gamma_{\mu}$ and $\gamma_{\rm e}$ are 
gyro-magnetic ratio of 
muon and electron, respectively, and ${\bf r}_{i}$ is the position vector of electron measured from the 
position of stopped muon in the crystal as shown in Fig.\ \ref{Fig:dipole} which shows the clock-wise circular 
motion (along the dashed circle) of the Cooper pairs in the plane perpendicular to the $z$-axis with  
nearly constant angular frequency $\omega_{i}\simeq\hbar/2m^{*}r_{i,{\perp}}^{2}$ for a 
certain $r_{i,\perp}$, with $m^{*}$ being the effective mass of the quasiparticles. 
Note that such a spin current of Cooper-pairs is induced 
(by the pair-breaking effect of $\mu^{+}$ as discussed in Ref.\ \citen{Miyake3} for the  
charge current of Cooper-pairs in the chiral superconducting state of Sr$_2$RuO$_4$) 
around the stopped $\mu^{+}$ at $r\lsim \xi$, with $\xi$ being the size of the Cooper pairs. 


Hereafter, we retain only the terms including 
${\hat \mu}_{\pm}\equiv {\hat \mu}_{x}\pm{\hat \mu}_{y}$ 
in Eq.\ (\ref{dipole1}) because we are interested in {the extra LFD  
$\Delta H$ induced by the circular motions of the Cooper pairs. 
Note that the spins of Cooper pairs are vanishing because $\uparrow\uparrow$ and 
$\downarrow\downarrow$ pairs gives no spin polarization while they induce the spin current in 
one-direction, breaking the helical-symmetry as shown in Fig.\ \ref{Fig:dipole}.   
Then, after straight forward calculations, we obtain
\begin{eqnarray}
& &
H_{\rm d}={\mu_{\rm B}}^{2}g_{\mu}g_{\rm e}\sum_{i}
\Biggl\{ -\frac{3\cos^{2}\theta_i}{4r^{3}}
\left[{\hat \mu}_{+}S_{i,+}e^{-2{\rm i}\phi_{i}(t)}+{\hat \mu}_{-}S_{i,-}e^{2{\rm i}\phi_{i}(t)}
\right]
\nonumber
\\
& &
\qquad\qquad\qquad\qquad\qquad\quad
-\frac{3\sin(2\theta_i)}{4r^{3}}
\left[{\hat \mu}_{+}e^{-{\rm i}\phi_{i}(t)}+{\hat \mu}_{-}e^{{\rm i}\phi_{i}(t)}\right]S_{i,z}
\Biggr\},
\label{dipole2} 
\end{eqnarray}
where $S_{i,\pm}\equiv S_{i,x}\pm S_{i,y}$, $\phi_{i}(t)\equiv \phi_{i}-\omega_{i} t$, and 
only terms, including $e^{\pm{\rm i}\phi_{i}(t)}$ and $e^{\pm 2{\rm i}\phi_{i}(t)}$ 
have been retained as discussed above.  Note, however, that the first term in Eq. (\ref{dipole2}) gives 
no contribution to $\Delta H$ because $S_{i,\pm}=0$ in the helical-symmetry broken 
superconducting state discussed in Sect.\ 3. 

Since the spin current of the Cooper pairs flows at around $r_{i}\sim \xi$, a fundamental 
magnetic field size ${\bar H}_{\mu}$, which the $\mu^{+}$ feels from each $S_{i,z}$ 
through the dipole-dipole interaction Eq.\ (\ref{dipole2}), is 
\begin{eqnarray}
{\bar H}_{\mu}\sim \frac{\mu_{\rm B}}{\xi^{3}}\langle S_{iz}\rangle,
\label{dipole3}
\end{eqnarray}
where $\langle S_{iz}\rangle$ is the $z$-component of the spin of the Cooper pairs at ${\bf r}_{i}$  
and is given roughly as  
\begin{eqnarray}
\langle S_{iz}\rangle\sim \frac{\Delta}{E_{\rm F}^{*}},
\label{dipole4}
\end{eqnarray}
where $\Delta$ and $E_{\rm F}^{*}$ are the superconducting gap and the effective Fermi energy of 
the quasiparticles. 
Considering a typical case $\xi\sim 10^{-7}\,$[m](=$10^{3}\,$[\AA]) and using 
$\mu_{\rm B}\simeq9.3\times10^{-24}\,$[JT$^{-1}$], 
${\bar H}_{\mu}/\langle S_{iz}\rangle\sim\mu_{\rm B}/\xi^{3}$ is estimated as 
\begin{eqnarray}
\frac{{\bar H}_{\mu}}{\langle S_{iz}\rangle}\sim 10^{-2}\,\,{\rm[T]}=10^{2}\,\,{\rm [G]}. 
\label{dipole4A}
\end{eqnarray}

The size of the extra local field distribution $\Delta H$ is roughly estimated by summing up 
the magnetic field contribution on the $\mu^{+}$ [given by Eq.\ (\ref{dipole2})] from the sites 
${\bf r}_{i}$ extending within the distance $\xi$ from a $\mu^{+}$ site, and by obtaining the 
mean-square root over the circulating period $T_{\rm period}\sim 4\pi m^{*}\xi^{2}/\hbar$ for 
$r\sim\xi$, which is estimated to be far smaller than the life-time of the $\mu^{+}$ of the 
order of $10^{-6}\,$[sec] because $T_{\rm period}\sim 10^{-8}\,$[sec] for 
$\xi\,$=$10^{-7}\,$[m](=$10^{3}\,$[\AA]) as above.  Note that the typical angular frequency 
of $\omega_{i}$ in $\phi_{i}(t)\equiv\phi_{i}-\omega_{i}t$ is given by $2\pi/T_{\rm period}$. 

The quantity in the bracket of the second term in the brace of Eq.\ (\ref{dipole2}) is equal to 
$2({\hat \mu}_{x}+{\hat \mu}_{y})\sin\phi_{i}(t)$ so that this term represents the existence of an 
oscillating magnetic field (with the angular frequency $\omega_{i}$) on the muon spin, 
$({\hat \mu}_{x}+{\hat \mu}_{y})$, from 
the Cooper pairs whose center is located at ${\bf r}_{i}$. At fixed time $t$, the length 
$r_{i}\equiv |{\bf r}_{i}|$, 
and the polar angle $\theta_{i}$, the summation with respect to 
$\phi_{i}$ around the circular orbit gives the magnetic field of the order of 
${\bar H}_{\mu}$ [Eq.\ (\ref{dipole3})]. 
This is because $\phi_{i}$ is distributed discretely on the circular orbit 
due to the atomic structure of ions. Then, at fixed $t$ and $r_{i}$, the summation with respect to 
$\theta_{i}$ gives a factor far smaller than 1 but non-vanishing in general, 
considering the fact that the distribution of the circular orbit 
is not symmetric (depending on the stopping position of $\mu^{+}$) and discrete as in that of $\phi_{i}$.  
Finally, the summation with respect to 
$r_{i,\perp}$ gives a factor $\xi/a$, with $a$ being the unit-cell size. Therefore, with the use of 
Eqs.\ (\ref{dipole2}) - (\ref{dipole4}), the magnetic field $H_{\mu}(t)$ acting on the $\mu^{+}$
spin component $({\hat \mu}_{x}+{\hat \mu}_{y})$ 
is roughly estimated as follows:
\begin{eqnarray}
H_{\mu}(t)\sim -3\bigr\langle\sin(2\theta_{i})\bigr\rangle\frac{{\bar H}_{\mu}}{\langle S_{iz}\rangle}
\frac{\Delta}{E_{\rm F}^{*}}\frac{\xi}{a}
\sin({\bar \omega}t),
\label{dipole5}
\end{eqnarray}
where $a$ is the lattice constant which is of the same order as the mean distance between 
quasiparticles, and ${\bar \omega}$ ($=2\pi/T_{\rm period}$) is the average angular frequency of 
circular motion of the Cooper pairs. 

The extra LFD $\Delta H$ in the superconducting state is obtained by taking 
the mean-square root average of $H_{\mu}(t)$ [Eq.\ (\ref{dipole5})] with respect to the circular motion as 
 \begin{eqnarray}
\Delta H\sim \frac{3}{2}\bigl|\bigl\langle\sin(2\theta_{i})\bigr\rangle\bigr|\frac{{\bar H}_{\mu}}{\langle S_{iz}\rangle}
\frac{\Delta}{E_{\rm F}^{*}}\frac{\xi}{a}. 
\label{dipole6}
\end{eqnarray}
With the use of Eq.\ (\ref{dipole4A}) and borrowing the relation of 
$\Delta(T=0)/E_{\rm F}^{*}\simeq(2/\pi^{2})[a/\xi(T=0)]$, valid in the s-wave pairing, 
the extra LFD $\Delta H(T=0)$ is estimated as 
\begin{eqnarray}
\Delta H(T=0)\sim \frac{3\times10^{2}}{\pi^{2}}\bigl|\bigl\langle\sin(2\theta_{i})\bigr\rangle\bigr|\,{\rm [G]}.
\label{dipole7}
\end{eqnarray}
This value of $\Delta H(T=0)$ is consistent with that observed in a series of compounds by the 
$\mu$SR measurement if $|\langle\sin(2\theta_{i})\rangle|$ is ${\cal O}(10^{-1})$, 
i.e., $\Delta H(T=0)\sim 1,$G.

}

\section*{Acknowledgments}
I have benefited much from conversations with J. Quintanilla, S. K. Ghosh, and J. F. Annett 
on theories for time-reversal symmetry broken states in a series of superconductors. 
This work is supported by JSPS KAKENHI Grant Number 17K05555, 
and EPSRC (UK) through the project ``Unconventional Superconductors: New paradigms for 
new materials'' (grant references EP/P00749X/1 and EP/P007392/1). 


\end{document}